\documentclass[12pt]{iopart}

\expandafter\let\csname equation*\endcsname\relax
\expandafter\let\csname endequation*\endcsname\relax
\usepackage[export]{adjustbox}
\usepackage{amsmath,amssymb,mathtools}
\DeclareMathOperator{\re}{Re}
\DeclareMathOperator{\im}{Im}
\DeclareMathOperator{\diag}{diag}
\usepackage{bm}
\usepackage{graphicx}
\usepackage{amsfonts}
\usepackage{dcolumn}
\usepackage{graphicx}
\usepackage{array}
\usepackage{tikz}
\usepackage{bm}
\usepackage[colorlinks=true,linkcolor=blue,citecolor=blue,urlcolor=blue]{hyperref}
\usepackage[utf8]{inputenc}
\usepackage[english]{babel}
\usepackage[T1]{fontenc}
\usepackage{array}
\usepackage[numbers]{natbib}
\usepackage[normalem]{ulem}

\newcommand{\delete}[1]{}
\newcommand{\newtext}[1]{#1}

\begin{document}

\title[New J. Phys. xx(2023) xxxxxx]{Resonance fluorescence of noisy systems}

\author{Rafał A. Bogaczewicz and Paweł Machnikowski$^*$}

\address{Institute of Theoretical Physics, Wrocław University of Science and Technology, Wybrzeże Wyspiańskiego 27, 50-370 Wrocław, Poland\\
$^*$Author to whom any correspondence should be addressed}
\ead{pawel.machnikowski@pwr.edu.pl}
\vspace{10pt}
\begin{indented}
\item[]xx August 2023
\end{indented}

\begin{abstract}
Light scattering from resonantly or nearly resonantly excited systems, known as resonance fluorescence, has been gaining importance as a versatile tool for investigating quantum states of matter and readout of quantum information, recently including also the inherently noisy solid state systems. In this work we develop a general theory of resonance fluorescence in the low excitation limit on systems in which the transition energy is subject to noise for two important classes of noise processes: white noise fluctuations that lead to phase diffusion and an arbitrary stationary Markovian noise process on a finite set of states. We apply the latter to the case of random telegraph noise and a sum of an arbitrary number of identical random telegraph noise contributions. We show that different classes of noise influence the resonance fluorescence spectrum in a characteristic way. Hence, the spectrum carries information on the characteristics of noise present in the physical system. 
\end{abstract}

%
%
%
%
%

\section{Introduction}

Light-matter interaction is one of the main tools for studying various properties of physical systems. In particular, resonant or nearly resonant light scattering, known as resonance fluorescence (RF) \cite{ScullyZubairy1997,Meystre2007} has been used for a long time to characterize systems of various kinds \cite{coope_nuclear_1977,wu_investigation_1975,diedrich_nonclassical_1987,kimble_photon_1977,Dunlap2019}. More recently, the RF technique has found a variety of applications in condensed matter systems, both in the physical investigation of quantum emitters, as well as in manipulating and reading out the quantum information encoded in solid-state qubits \cite{muller_resonance_2007,wrigge_efficient_2008}. It has been used to observe spin dynamics in semiconductor quantum dots (QDs) \cite{Vamivakas2010,Delteil2014}, to interface \cite{Ylmaz2010} and entangle \cite{Delley2017} QD spins with single photons, to generate indistinguishable photons \cite{Scholl2019}, to read out spin states in QDs \cite{Lu2010,Vamivakas2009,Ylmaz2010,Delteil2014,Vamivakas2010} and silicon defects \cite{Broome2017a} as well as to demonstrate quantum-optical effects in macroscopic superconducting qubits \cite{Astafiev2010,toyli_resonance_2016}. The recent observation of RF from a waveguide-coupled solid-state emitter \cite{Errando-Herranz2021} opens a perspective of on-chip device integration. 

\newtext{The resonance fluorescence from a single unperturbed quantum emitter shows different properties under weak and strong excitation. In the former case, energy conservation for each single-photon scattering event leads to a single line with the broadening limited by the laser line width \cite{ScullyZubairy1997}. Under strong excitation, the modulation of the system state due to Rabi rotations gives rise to a triplet of broadened lines separated by the Rabi frequency, referred to as the Mollow triplet \cite{Mollow1969,ScullyZubairy1997}.}

No physical system is completely isolated from its environment. The environmental impact is of particular importance in solid-state systems, where the optical properties are to a large extent influenced by the coupling to charge \cite{Matthiesen2015,Preuss:22} or spin fluctuations (both nuclear \cite{Malein2016} and electronic \cite{Kasprzak2022}), as well as to lattice vibrations. The latter can be induced in a coherent way, leading to controllable modification of the scattering spectrum \cite{Wigger2021,Weiss2021}, but in most cases is a source of noise \cite{Axt2005,Iles-Smith2017}. When treated as a classical background noise, the environmental fluctuations are often modeled using Gaussian distributions \cite{Matthiesen2015,Preuss:22}. However, it is generally believed that the underlying physics involves discrete state changes of nearby physical objects, like nuclear or dopant spin flips, or charging and discharging of defects, and such dynamics is indeed observed in certain experiments \cite{Kasprzak2022}. Therefore, a more fundamental model for the description of noise needs to be based on telegraph-noise dynamics. 

While numerous studies considered fluctuations in the phase  \cite{Burshtein1988,Kofman1990,Zoller1977,Georges1981,Eberly1984}, amplitude  \cite{Burshtein1988,Avan1977,Chaturvedi1981}, and  frequency \cite{Burshtein1988,Zoller1977,Rangwala1990} of the laser beam, much less attention has been devoted to the effect of the environmental noise on RF. The existing studies include the particular case of interaction with phonons  \cite{Ahn2005,Flagg2009} and the dynamics of inversion \cite{Vemuri2001}, coherence \cite{Cai2020}, entanglement \cite{Ke2005}, as well as non-linear wave-mixing response \cite{Green2001} in two-level systems subject to environmental fluctuations. Apart from studying the detrimental effects of noise on the dynamics of quantum systems, the latter can also be used as noise sensors in order to determine the properties of the noise itself. Such noise characterization is crucial for the robustness of quantum systems, hence considerable effort has recently been invested in the development of noise spectroscopy techniques \cite{sakuldee2020,Farfurnik2023,Sung2021,barr2022}.

In this work, we generalize the recently proposed description of RF from a deterministically modulated two-level system \cite{Wigger2021,Weiss2021} to systems subject to random fluctuations. We formulate a theory of low-excitation RF (coherent Rayleigh scattering) spectra for a system with the transition energy subject\delete{ed} to white noise or an arbitrary Markovian random process that shifts the transition energy between a number of discrete spectral positions. In particular, we consider single-source symmetric and asymmetric telegraph noise and multi-source telegraph noise.

By relating the scattering spectrum to the formal characteristics of the underlying noise process, we are able to show that the spectrum bears clear fingerprints of the properties of noise: In the white noise (phase diffusion) case, the noise leads to Lorentzian line broadening. In contrast, a slow discrete process on a small set of states results in a multiple of Lorentzian lines that merge into a broad Gaussian feature when the number of process states increases so that they become dense on the energy axis. This picture changes when the process is fast. In this case a motional narrowing effect leads to the appearance of a single Lorentzian line. 

The paper is organized as follows. In Sec.~\ref{sec:Model} we describe the system and define its general model. Sec.~\ref{sec:theory-general} contains the essential definitions and presents the general framework of our theoretical description; here we also present the theory for the simplest case of white noise. Sec.~\ref{sec:theory-markov} contains the central formal result of the paper: the theory of the resonance fluorescence spectrum for a Markovian noise process. Sec.~\ref{sec:results} presents the results obtained by applying the theory to selected noise processes. The paper is concluded by Sec.~\ref{sec:conclusions}.

\section{System and Model}
\label{sec:Model}

We consider light scattering on a two-level system which is subject to environmental fluctuations that randomly shift the energy of the excited state. As in the standard model of resonance fluorescence \cite{ScullyZubairy1997}, the system is driven by a resonant or nearly resonant monochromatic laser light and undergoes spontaneous emission. 

We denote the laser frequency by $\omega_L$ and the system states by $\arrowvert 0 \rangle$ and $\arrowvert 1 \rangle$. Let $\hbar\omega_0(t)$ be the time-dependent (fluctuating) energy difference between these states. The system is then described by the Hamiltonian
\begin{equation*}
H(t) = \hbar\omega_0(t)\arrowvert 1 \rangle \langle 1 \arrowvert - \bm{d}\cdot\bm{E}(t),
\end{equation*}
where 
$\bm{E}(t) = (1/2)\left(\bm{E_0}e^{-i\omega_L t} + \text{c.c.}\right) $
is the laser field (treated classically) with the amplitude $\bm{E}_0$ and $\bm{d}$ is the dipole moment operator. We assume $\langle 0 \arrowvert \bm{d} \arrowvert 0 \rangle = \langle 1 \arrowvert \bm{d} \arrowvert 1 \rangle = \bm{0}$. 
The system relaxation due to spontaneous emission is accounted for by the Lindblad dissipator
\begin{equation*}
L\left[\rho(t)\right]=\gamma\left(\sigma_-\rho(t)\sigma_+-\frac{1}{2}\left\{\sigma_+\sigma_-,\rho(t)\right\}\right),
\end{equation*}
where $\rho$ is the density matrix of the system, $\left\{A,B\right\}=AB+BA$, and $\sigma_+=\sigma_-^\dagger = \arrowvert 1 \rangle \langle 0 \arrowvert$.
The system state in the rotating frame is defined by
\begin{equation}
\tilde{\rho}(t)=e^{i\omega_L t \arrowvert 1\rangle\langle 1\arrowvert}\rho(t)e^{-i\omega_L t \arrowvert 1\rangle\langle 1\arrowvert}. 
\label{eq:rotating}
\end{equation}
The Hamiltonian in this rotating frame and in the rotating wave approximation is
\begin{equation*}
\tilde{H}(t)=-\hbar\Delta(t)\arrowvert 1 \rangle \langle 1 \arrowvert -\frac{\hbar\Omega}{2}\left(\sigma_-+\sigma_+\right),
\end{equation*}
where we define the detuning $\Delta(t) = \omega_L-\omega_0(t)$ and the Rabi frequency 
$\Omega=(\bm{E}_0^*/\hbar) \cdot \langle 0 \arrowvert \bm{d} \arrowvert 1 \rangle$
that we assume real. 
The Master Equation describing the evolution of the system state has the form 
\begin{equation}
\frac{d\tilde{\rho}(t)}{dt} = -\frac{i}{\hbar}\left[\tilde{H}(t),\tilde{\rho}(t)\right] + L\left[\tilde{\rho}(t)\right]. \label{eq:ewolucja_ME}
\end{equation}

The randomly changing detuning $\Delta(t)$, which reflects the environmental fluctuations, is the central feature of our model. Formally, it is described by a stochastic process, the properties of which may depend on the physical situation. Here we assume that $\Delta(t)$ is a stationary Markov process with a unique stationary probability distribution $p^{(\mathrm{st})}$ reached asymptotically in the long time limit. Two particular examples will be discussed in the following.

\section{The resonance fluorescence spectrum of a noisy system}
\label{sec:theory-general}

In this section we define the formal quantities relevant to the RF spectrum, present the general framework of the theoretical model and discuss the simplest case of phase diffusion due to white noise. 

In the Markov approximation, the detected spectrum of scattered light can be related to the system autocorrelation function $G(t, t+\tau) = \langle\sigma_+(t)\sigma_-(t+\tau)\rangle$ by \cite{ScullyZubairy1997}
\begin{equation}
F(\omega) = \re\int_0^{\infty}d\tau e^{i(\omega-\omega_L)\tau}e^{-\Gamma \tau} G(t, t+\tau). \label{eq:widmo_RF}
\end{equation}
Here $\sigma_{\pm}(t)$ are operators in the Heisenberg picture relative to the rotating-frame evolution as defined by Eq.~(\ref{eq:ewolucja_ME}) and  $\Gamma$ is the (Lorentzian) instrumental broadening accounting for the finite resolution of the detection. 

Let us formally denote the solution of Eq.~(\ref{eq:ewolucja_ME}) by $\tilde{\rho}(t) = \mathfrak{L}_{t_0,t} \left[\tilde{\rho}(t_0)\right]$, where $\mathfrak{L}_{t_1,t_2}$ is the evolution superoperator. The Lax quantum regression theorem then yields the autocorrelation function in the form \cite{Meystre2007,Weiss2021}
\begin{equation}
\label{eq:autokorelacja_Lax}
G(t, t+\tau) = \Tr\left(\sigma_-\mathfrak{L}_{t,t+\tau}\left[\mathfrak{L}_{t_0,t}\left[\tilde{\rho}(t_0)\right]\sigma_+\right]\right).
\end{equation}
Here $t_0$ is the initial moment of the evolution, while $t-t_0$ is a sufficiently long time for the system to reach its steady state.

The total scattering intensity is $I_{\text{tot}} = \int_{-\infty}^{\infty}F(\omega) d\omega$. In the noise-free limit, the RF spectrum consists of a Dirac delta (broadened by the instrumental resolution) corresponding to elastic light scattering, which survives to various extent in the noisy case. Its intensity will be denoted by $I_{\text{el}}$. The remaining part of the spectrum is due to inelastic scattering induced by the random fluctuations. Its intensity is $I_{\text{inel}}$.

\newtext{
The equations of motion for the elements $\rho_{01}$, $\rho_{10}$, $\rho_{11}$ of the density matrix, following from Eq.~(\ref{eq:ewolucja_ME}), have the form $\dot{\rho}_{jl} = a_{jl}\rho_{jl} + i\Omega\sum_{mn}b_{jl,mn}\rho_{mn}$, where $a_{11}=-\gamma$, $a_{01}=a_{10}^*=i\Delta-\gamma/2$, and 
$b_{11,10}=-b_{11,01}=b_{10,11}=-b_{01,11}=1/2$. The same holds for an arbitrary matrix, not necessarily a density matrix. Since Eq.~(\ref{eq:ewolucja_ME}) is trace-preserving, one has $\rho_{00} = c_0 - \rho_{11}$, where $c_0$ is a constant determined by the initial values ($c_0=1$ for a density matrix). In the absence of the laser field ($\Omega=0$) the equation of motion can be solved trivially to yield the zeroth-order propagation 
\begin{equation*}
\rho_{jl}^{(0)}(t)=\left[\mathfrak{L}^{(0)}_{t_0,t}\rho(t_0)\right]_{jl}
= e^{\int_{t_0}^{t}ds a_{jl}(s)}\rho_{jl}(t_0).
\end{equation*}
}
In the weak excitation regime, one can \newtext{then} solve Eq.~(\ref{eq:ewolucja_ME}) iteratively in the subsequent orders $r>0$ in $\Omega$,
\newtext{
\begin{equation*}
\rho_{jl}^{(r)}(t)=\left[\mathfrak{L}^{(r)}_{t_0,t}\rho(t_0)\right]_{jl}
= i\Omega \int_{t_0}^t ds e^{\int_{s}^{t}ds' a_{jl}(s')}\sum_{mn}b_{jl,mn}\rho_{mn}^{(r-1)}(s).
\end{equation*}
These equations fully define the perturbative expansion of the evolution superoperator $\mathfrak{L}_{t_1,t_2}$ in powers of $\Omega$.
Substituting this evolution into Eq.~\eqref{eq:autokorelacja_Lax} one finds, 
}
in the leading order of $\Omega^2$, the autocorrelation function for an arbitrary time-dependent energy shift $\Delta(t)$ \delete{has} \newtext{in} the form \delete{[26,27]}
\begin{equation*} 
\hbox{\delete{$G(t, t+\tau) = \frac{\Omega^2}{4}\int_{-\infty}^{t}ds e^{-\frac{\gamma}{2}(t-s)}e^{-i\Phi(t,s)} \int_{-\infty}^{t+\tau}ds'e^{-\frac{\gamma}{2}(t+\tau-s')}e^{i\Phi(t+\tau,s')}$,\quad\quad (5)}} 
\end{equation*}
\delete{where}
\begin{equation*}
\hbox{\delete{
$\Phi(t_b,t_a)=\int_{t_a}^{t_b}dt\Delta(t)$.}} 
\end{equation*}
\delete{
When $\Delta$ is a random process, it is convenient to rewrite Eq.~(5) in a time-ordered form, such that the exponential terms have the form $e^{i\Phi(t_d, t_c)\pm i\Phi(t_b, t_a)}$ with $t_a<t_b<t_c<t_d$. This is achieved by changing the variables to $s=t+u$ and $s'=t+u'$, splitting the second integral into segments $(-\infty,0)$ and $(0,\tau)$ and then dividing the first part into contributions with $u>u'$ and $u<u'$. After these transformations one gets
}
\begin{align} \label{eq:autokorelacja_2}
\MoveEqLeft G(t, t+\tau)  = 
\frac{\Omega^2}{4}e^{-\frac{\gamma}{2}\tau}\int_{-\infty}^{0} du \left[ \int_{0}^{\tau} du' e^{\frac{\gamma}{2}(u+u')}
\overline{e^{i\Phi(\tau,u')-i\Phi(0,u)}} \right .\\
& {}\quad \hspace{3.0cm}+ \int_{-\infty}^{u} du' e^{\frac{\gamma}{2}(u+u')}
\overline{e^{i\Phi(\tau,0)-i\Phi(u,u')}} + \left.\int_{-\infty}^{u} du' e^{\frac{\gamma}{2}(u+u')}
\overline{e^{i\Phi(\tau,0)+i\Phi(u,u')}}\right], \nonumber
\end{align}
where
\newtext{
\begin{equation}
\Phi(t_b,t_a)=\int_{t_a}^{t_b}ds\Delta(s). \label{eq:Phi}
\end{equation} }
\newtext{Here we changed the variables according to $s=t+u$, set $t-t_0\to\infty$ (steady-state regime),} averaged over the realizations of the noise (denoted by a line above the averaged quantities) and used the fact that the noise is stationary, hence 
\begin{equation*}
\overline{e^{i\Phi(t_d+s, t_c+s)\pm i\Phi(t_b+s, t_a+s)}} = 
\overline{e^{i\Phi(t_d, t_c)\pm i\Phi(t_b, t_a)}}.
\end{equation*}
\newtext{
The detailed derivation along with a graphical interpretation of the evolution paths contributing to the RF signal can be found in the Supplementary Material to Ref.~\cite{Weiss2021}.
}

Before developing the theory for an arbitrary Markovian noise process on a discrete state space, we find the explicit form of the correlation function in the case of simple phase diffusion. We assume that 
\begin{equation}
\Delta(t) = \Delta_{WN}(t) + \overline{\Delta}, \label{eq:odstrojenie_nieskorelowany_szum}
\end{equation}
where $\Delta_{WN}(t)$ is a stationary white noise process with $\langle \Delta_{WN}(t) \rangle = 0$ and $\langle \Delta_{WN}(t)\Delta_{WN}(t+\tau) \rangle = D \delta(\tau)$. Then, the phase shift given in Eq.~(\ref{eq:Phi}) has a normal distribution with the mean value $\overline{\Delta}(t_b-t_a)$ and variance $D(t_b-t_a)$, hence $D$ is the phase diffusion coefficient. With this Gaussian distribution, Eq.~(\ref{eq:autokorelacja_2}) can easily be evaluated by using the statistical independence of phase shifts over non-overlapping periods of time. This yields
\begin{equation}
G(t, t+\tau) = \frac{\Omega^2\left(1+\frac{D}{\gamma}e^{[i\overline{\Delta}-(\gamma+D)/2]\tau}\right)}{(\gamma+D)^2+4\overline{\Delta}^2}. \label{eq:funkcja_autokorelacji_szum_bialy}
\end{equation}
The RF spectrum obtained by substituting Eq.~(\ref{eq:funkcja_autokorelacji_szum_bialy}) to Eq.~(\ref{eq:widmo_RF}) is a sum of elastic and inelastic peaks that are, respectively, given by
\begin{equation} \label{eq:F_el}
F_{\mathrm{el}}(\omega) = \frac{\Omega^2}{(\gamma + D)^2+4\overline{\Delta}^2}\frac{\Gamma}{(\omega-\omega_L)^2+\Gamma^2} 
\end{equation}
and
\begin{align}\label{eq:F_inel}
F_{\mathrm{inel}}(\omega) = & \frac{\Omega^2}{(\gamma + D)^2+4\overline{\Delta}^2}\frac{D}{\gamma} \frac{(\gamma+D)/2+\Gamma}{(\omega+\overline{\Delta}-\omega_L)^2+\left[(\gamma+D)/2+\Gamma\right]^2}.
\end{align}
Thus, the elastic scattering line is located at the laser frequency, while the inelastic line appears at the average system transition frequency.

By integrating we find the total intensity 
\begin{equation}
I_{\mathrm{tot}} = I_0 \frac{1+D/\gamma}{\left(1 + D/\gamma\right)^2+\left(2\overline{\Delta}/\gamma\right)^2}, \label{eq:I_tot_dyfuzja}
\end{equation}
as well as the intensities of the elastic and inelastic components
\begin{equation}
I_{\mathrm{el}} = I_0 \frac{1}{\left(1 + D/\gamma\right)^2+\left(2\overline{\Delta}/\gamma\right)^2}, \quad I_{\mathrm{inel}} = \frac{D}{\gamma}I_{\mathrm{el}}. \label{eq:I_el_dyfuzja}
\end{equation}
Here and in the following we relate the scattering intensities to the standard resonance fluorescence intensity in weak excitation limit \cite{ScullyZubairy1997}, $I_0 = \pi\Omega^2/\gamma^2$.

\section{Theory of the RF spectrum for arbitrary Markovian noise}
\label{sec:theory-markov}

In this section we develop the theory of the resonance fluorescence spectrum for a system subject to noise that can be described by a continuous-time stationary Markov process on a finite set of states $\{\Delta_i\}_{i=1}^k$. This model can also be used as an approximation to more general Markov processes, based on a physically motivated truncation and discretization of the state space of the noise process. 

The process is characterized by the transition probabilities $P_{m,n}(\tau)=\mathcal{P}[\Delta(t+\tau)=\Delta_m | \Delta(t)=\Delta_n]$ (with $\mathcal{P}$ denoting the conditional probability) forming the transition matrix $P(\tau)=\exp(C\tau)$, with the generator $C = dP(\tau)/d\tau|_{\tau=0}$. As shown in \ref{app:average}, for such a process
\begin{align}\label{eq:average}
\MoveEqLeft \overline{e^{i\Phi(t_d, t_c)\pm i\Phi(t_b, t_a)}} = \Tr P_\infty e^{B\left(t_d-t_c\right)}
P_{t_c-t_b}e^{B^{(*)}\left(t_b-t_a\right)},
\end{align}
where 
$B = C+i\diag(\Delta_1, \Delta_2, \ldots, \Delta_k)$, $P_\infty = \lim_{\tau\to\infty} P_\tau = p^{(\mathrm{st})}( 1, \ldots,1)$, with $p^{(\mathrm{st})}$ representing the column vector of stationary probabilities, and the conjugation in the last term refers to the '$-$' sign on the left-hand side.

Substituting Eq.~\eqref{eq:average} to Eq.~\eqref{eq:autokorelacja_2} one gets
\begin{align} \label{eq:correl-general}
G(t, t+\tau) = & \frac{\Omega^2}{4}\Tr P_\infty 
\int_0^\tau du' e^{\left(B-\frac{\gamma}{2}\right)(\tau-u')} P_{u'} \int_{-\infty}^{0} du e^{(\frac{\gamma}{2}-B^*)u} \\
& + \frac{\Omega^2}{4}\Tr P_\infty e^{\left( B-\frac{\gamma}{2} \right)\tau}
\int_{-\infty}^{0} du e^{\frac{\gamma}{2}u}P_{-u} 2\re \int_{-\infty}^{u} du' e^{\frac{\gamma}{2}u'+B(u-u')}. \nonumber
\end{align}
Upon substituting to Eq.~\eqref{eq:widmo_RF}, the first term in Eq.~(\ref{eq:correl-general}) can be factorized by reordering the integrals with respect to $\tau$ and $u'$ and then changing the variable according to $\tau=u'+s$. The second term can be evaluated directly. As a result one gets
\begin{align}\label{eq:widmo_RF_TN_do_wykresu}
F(\omega) = & \frac{\Omega^2}{4} \re \Tr P_\infty 
\int_0^\infty ds e^{\left(i(\omega-\omega_L) -\Gamma +B-\frac{\gamma}{2}\right)s}  \\
&\quad\times \left[
\int_0^\infty du e^{\left(i(\omega-\omega_L)-\Gamma+C\right)u}
\int_0^\infty du' e^{\left(B^*-\frac{\gamma}{2}\right)u'} + \int_0^\infty du e^{\left(C-\gamma \right)u} 2\re\int_0^\infty du' e^{\left(B-\frac{\gamma}{2}\right)u'}\right] \nonumber \\
= & - \frac{\Omega^2}{4}\re\Tr P_\infty
\left( i(\omega-\omega_L)-\Gamma+B-\frac{\gamma}{2} \right)^{-1} \nonumber\\
&\quad\times \left[ \left( i(\omega-\omega_L)-\Gamma+C \right)^{-1}
\left( B^*-\frac{\gamma}{2} \right)^{-1} + 2\left(C-\gamma \right)^{-1} 
\re \left( B-\frac{\gamma}{2} \right)^{-1} \right]. \nonumber
\end{align}
While this closed analytical form may be convenient for evaluating the spectrum in the case of a small state space of the process, much more insight is gained by relating the RF spectrum to the spectral properties of the generator $C$. To this end, we transform $C$ to the Jordan form (over the field of complex numbers) by the similarity transformation
\begin{equation*}
S^{-1}CS = \bigoplus_j C_j,
\end{equation*}
where $C_j$ are Jordan blocks belonging to the respective eigenvalues $\lambda_j$ of algebraic multiplicity $d_j$. We then apply the Jordan-Chevalley decomposition 
$C_j = \lambda_j\mathbb{I}^{(d_j)}+\mathbb{N}_j$, 
where $\mathbb{N}_j$ is a nilpotent $d_j$-dimensional matrix, $\left(\mathbb{N}_j\right)_{kl}=\delta_{k,l-1}$, and $\mathbb{I}^{(d)}$ is the $d$-dimensional unit matrix.
Let $\Pi_j$ be the projector on the subspace supporting $C_j$. 
Since 
\begin{equation*}
e^{C_j\tau} = e^{\lambda_j \tau}\sum_{n=0}^{d_j-1}\frac{\left(\mathbb{N}_j\right)^n}{n!}\tau^n,
\end{equation*}
one finds 
\begin{equation*}
P_\tau = e^{C\tau} = \sum_j e^{\lambda_j \tau}\sum_{n=0}^{d_j-1}P_{jn}\tau^n
\end{equation*}
with
\begin{equation*}
P_{jn} = S\Pi_j \frac{\left(\mathbb{N}_j\right)^n}{n!} \Pi_j S^{-1}.
\end{equation*}
In the case of a diagonalizable matrix $C$, $d_j = 1$ for all $j$, the above procedure reduces to simple diagonalization, and $\lambda_j$ become eigenvalues of $C$ in the most common sense.

Using this result in the first term of Eq.~\eqref{eq:correl-general}, substituting to Eq.~\eqref{eq:widmo_RF}, and performing the integrations one finds
\begin{align}
F(\omega) = & \frac{\Omega^2}{4}\re \Tr P_\infty \Biggl\{\sum_j \sum_{n=0}^{d_j-1}  \frac{d^n}{d\lambda_j^n}  \left(i(\omega-\omega_L)-\Gamma+\lambda_j\right)^{-1}\left( \frac{\gamma}{2}-B +\lambda_j \right)^{-1}P_{jn}\left(B^* - \frac{\gamma}{2}\right)^{-1}  \nonumber \\
& +  \left(i(\omega-\omega_L)+B-\Gamma-\frac{\gamma}{2}\right)^{-1} \label{eq:widmo_RF_TN} \\
&\times \left[\sum_j \sum_{n=0}^{d_j-1} \frac{d^n}{d\lambda_j^n} \left( B-\frac{\gamma}{2} -\lambda_j \right)^{-1}P_{jn}\left(B^* - \frac{\gamma}{2}\right)^{-1} +  \left(C-\gamma\right)^{-1}2\re\left(\frac{\gamma}{2}-B\right)^{-1}
\right]\Biggr\}. \nonumber
\end{align}
The essential factors that describe the form of the spectrum are those depending on the frequency $\omega$.

The first term defines Lorentzian features and more general line shapes in the case of degenerate eigenvalues ($d_j>1$), as well as the corresponding dispersive contributions, with central frequencies at $\omega_\mathrm{L}+\im\lambda_j$ and broadened by $-\re\lambda_j$ (here additionally broadened by the instrumental resolution $\Gamma$). The existence of a stationary state implies that one of the eigenvalues (say, $\lambda_0$) is zero and $P_{00}=P_{\infty}$. This contribution leads to an unbroadened (apart from the instrumental broadening) elastic scattering peak centered at $\omega=\omega_L$. For $\lambda_0$ there is no dispersive function. Each real non-zero eigenvalue leads to a broadened spectral feature at $\omega=\omega_L$. 

The second term is a sum of simple Lorentzians and the corresponding dispersive contributions with positions and widths determined by the spectrum of $B$ and further broadened due to spontaneous emission. 

The total and elastic scattering intensities, found by integrating Eq.~(\ref{eq:widmo_RF_TN}), are respectively
\begin{equation}
I_{\mathrm{tot}} = -\frac{I_0 \gamma}{2}\Tr P_{\infty}\re\left(B-\frac{\gamma}{2}\right)^{-1} \label{eq:I_tot_ogolnie}
\end{equation}
and
\begin{equation}
I_{\mathrm{el}} = \frac{I_0\gamma^2}{4} \re\Tr P_\infty \left(B-\frac{\gamma}{2}\right)^{-1} P_\infty\left(B^*-\frac{\gamma}{2}\right)^{-1}.\label{eq:I_el_ogolnie}
\end{equation}

As an application of this formalism, we study in detail the special case of $N$ identical noise sources, each generating telegraph noise (TN).

A single noise source has two states that contribute $\Delta = \pm\frac{\Delta_0}{2\sqrt{N}}$ to the system energy shift. The switching rates between the two states of the noise are $\beta_{\uparrow}$ and $\beta_{\downarrow}$, leading to stationary probabilities $\beta_\downarrow / \left(\beta_\uparrow + \beta_\downarrow\right)$ and $\beta_\uparrow / \left(\beta_\uparrow + \beta_\downarrow\right)$ for the two noise states. 
For $N$ identical and additively contributing noise sources the space of possible noise states is composed of $N+1$ values of the total system detuning,
\begin{equation}
\Delta_{j} = \overline{\Delta} - \sqrt{N}\frac{\Delta_0}{2} + \frac{j}{\sqrt{N}}\Delta_0, \quad j=0,\ldots,N, \label{eq:Delta_j}
\end{equation}
corresponding to $j$ sources in the ``upper state''.
Here $\overline{\Delta}$ is the mean detuning (the laser detuning from the noise-free transition energy) and the renormalization by a factor $1/\sqrt{N}$ assures convergence in the limit of $N\to\infty$.

For this case the generator $C$ is a $N+1$-dimensional tridiagonal matrix, where the non-zero elements are
\begin{align}
C_{j,j} &=  (j-N)\beta_\uparrow -j\beta_\downarrow,  &j = 0,\ldots, N, \nonumber \\
C_{j-1,j} &= j\beta_\downarrow,  &j = 1,\ldots, N, \label{eq:macierz_C} \\
C_{j,j-1} &= (N+1-j)\beta_\uparrow,  &j = 1,\ldots, N. \nonumber
\end{align}
The stationary probability follows the binomial distribution
\begin{equation}
p_j^{(\mathrm{st})} = \binom{N}{j}
\frac{\beta_\uparrow^j \beta_\downarrow^{N-j}}{(\beta_\uparrow+\beta_\downarrow)^N}. \label{eq:binom}
\end{equation}

\section{Results}
\label{sec:results}

In this section we present the results for the RF spectrum based on the theory developed in Sec.~\ref{sec:theory-general} and Sec.~\ref{sec:theory-markov} for three noise models. For the white noise calculations we set $\Gamma / \gamma=0.25$. For the discrete process we choose $\gamma = 4\Gamma = 0.02\Delta_0$.
\newtext{As a natural unit for presenting and comparing the RF spectra we will use the maximum value of the spectrum for an unperturbed system under weak resonant excitation \cite{ScullyZubairy1997}, $F_0 = \Omega^2/(\gamma^2 \Gamma)$, which can be obtained from Eq.~(\ref{eq:F_el}) with $D=\overline{\Delta}=0$ and $\omega=\omega_{\mathrm{L}}$. All the spectra presented in the following will be related to this quantity.}

\subsection{White noise}

\begin{figure}[tb]
\includegraphics[width=0.63\linewidth,center]{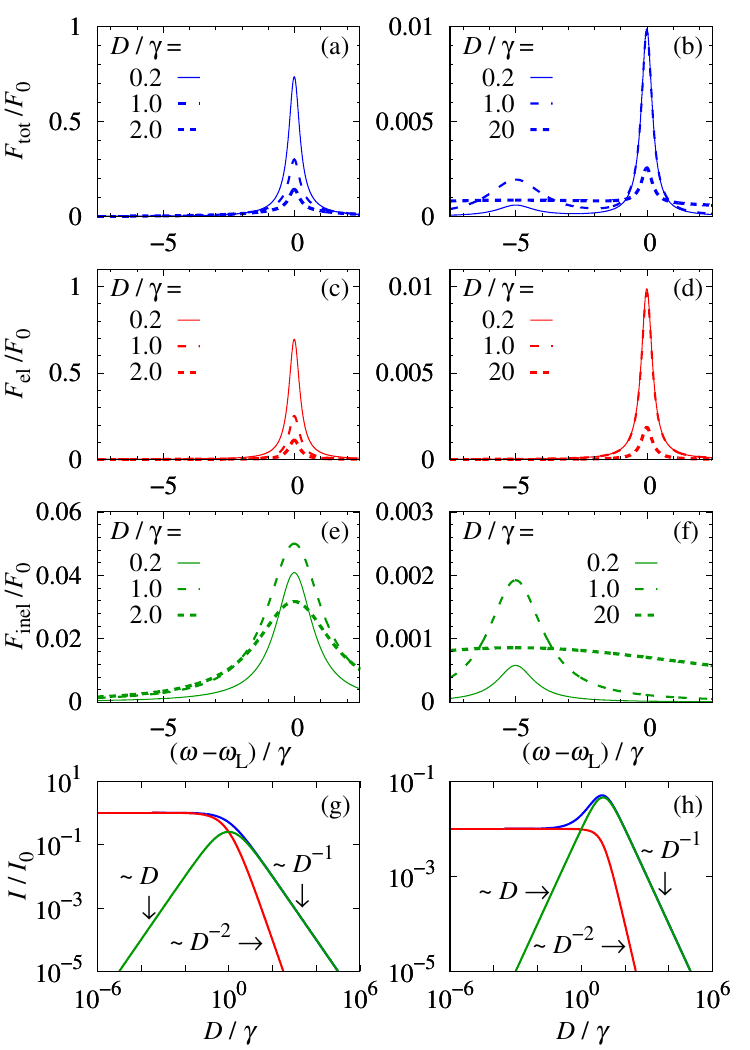}
\caption{\label{fig:dyfuzja} The RF spectra and the scattering intensities for uncorrelated noise under resonant excitation (left column) and detuned excitation with $\overline{\Delta}/\gamma=5$ (right column). (a,b) Total spectrum; (c,d) elastic contribution; (e,f) inelastic contribution; (g,h) scattering intensities as a function of the phase diffusion coefficient, with line colors corresponding to (a-f).} 
\end{figure}

We start the discussion with the case of a system affected by white noise. Fig.~\ref{fig:dyfuzja} shows the results for this case, calculated from Eq.~(\ref{eq:F_el}) and Eq.~(\ref{eq:F_inel}), under resonant and detuned excitation in the left and right columns, respectively. In both cases, the \delete{the} RF spectrum is composed of two Lorentzians. In the resonant case they overlap, while for a detuned excitation they are split. As expected, the overall intensity is also lower in the latter case.

The total spectrum, shown in Fig.~\ref{fig:dyfuzja}(a,b), is decomposed into the elastic and inelastic contributions in Fig.~\ref{fig:dyfuzja}(c,d) and Fig.~\ref{fig:dyfuzja}(e,f), respectively. While the positions of the spectral features do not change, the evolution of their intensities and of the width of the non-elastic line are clearly visible. The width of the inelastic contribution grows with $D$. The intensity of the elastic contribution starts to decrease when $D \sim \gamma$, while the intensity of the inelastic one changes non-monotonically with $D$, reaching a maximum for $D \sim \gamma$. The decrease of the inelastic contribution for weak noise is an obvious consequence of restoring the noise-free limit of purely elastic scattering. For strong noise both components decrease because the increasing spread of the transition energy reduces the effective overlap with the excitation frequency, which affects the excitation intensity. The appearance of an additional broadened spectral line at a spectral position bound to the transition energy in addition to the  elastic line at the laser position can be understood by invoking the model of a classical charged harmonic oscillator driven by a periodic force. This analogy is formally validated by the fact that in the leading order in $\Omega$, the whole emitted light is coherent, that is, originating from the transition dipole induced coherently by the laser field. In its steady state, the classical system oscillates periodically with the laser frequency, which gives rise to a sharp line at this spectral position. However, any perturbation of the steady-state evolution leads to the appearance of a damped transient at the eigenfrequency of the system renormalized by damping. Here the noise serves as a perturbation that permanently excites the transient oscillations and simultaneously damps the coherence due to phase diffusion, leading to a broadened line.

A systematic study of the intensities as a function of the noise strength $D$ is presented in Fig.~\ref{fig:dyfuzja}(g,h). As can also be seen directly from Eq.~(\ref{eq:I_el_dyfuzja}), the dependence has asymptotically a power-law character. In the case of resonant excitation, $\overline{\Delta}=0$, the absence of noise ($D\to 0$) leads to permanent resonance condition, maximizing the total scattering intensity. For a sufficiently large detuning, $\overline{\Delta}>\gamma/2$, the interplay of the decreasing elastic scattering and non-monotonic inelastic one leads to the appearance of a maximum of the total scattering intensity at $D = 2\overline{\Delta}-\gamma$. In this case one observes a noise-induced enhancement of scattering: At the maximum, the total intensity is larger than in the noise-free case by a factor of $(\gamma^2+4\overline{\Delta}^2)/(4\gamma\overline{\Delta})$.  

\subsection{Single-source telegraph noise (TN)}

In this section we discuss the results for the scattering spectra and line intensities for a single source of random telegraph noise, depending on the characteristics of the noise dynamics (the switching rates $\beta_{\uparrow,\downarrow}$) and laser detuning $\overline{\Delta}$ from the average transition energy. 
In this case, the detuning takes randomly two values, hence the set of states of the stochastic process is reduced to  $\Delta=\overline{\Delta}\pm\Delta_0/2$. 
The generator in Eq.~(\ref{eq:macierz_C}) reduces to
\begin{equation}
C = \left(\begin{array}{cc}
-\beta_{\uparrow} &  \beta_{\downarrow} \\
 \beta_{\uparrow} & -\beta_{\downarrow} \\ \end{array}\right). \label{eq:C_2x2}
\end{equation}
Its eigenvalues are $\lambda_0 = 0$ and $\lambda_1 = -\left(\beta_{\uparrow}+\beta_{\downarrow}\right)$. Both eigenvalues are non-degenerate, hence the spectrum is composed of four simple Lorentzians and, possibly, the corresponding dispersive functions.
 For the sake of presentation we set $\beta_\uparrow = \beta(1-x)$ and $\beta_\downarrow = \beta (1+x)$, where $x\in [-1,1]$.
We begin with the case of symmetric noise ($x=0$, that is, $\beta_{\uparrow}=\beta_{\downarrow}=\beta$) and then describe the effects of noise asymmetry. 

\subsubsection{\textbf{Symmetric switching}}

\begin{figure}[tb]
\includegraphics[width=0.6\linewidth,center]{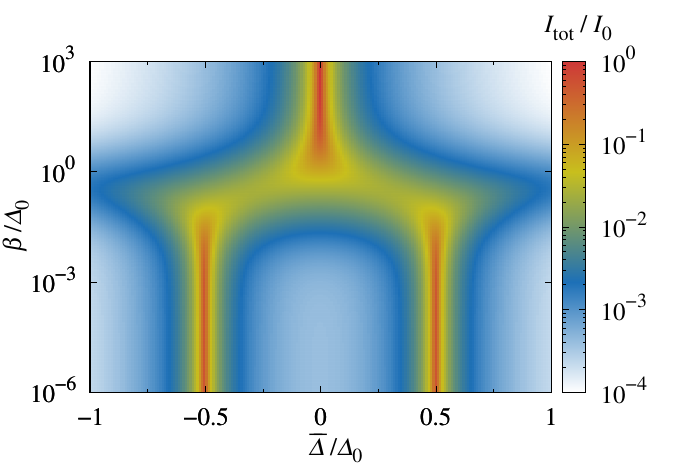}
\caption{\label{fig:mapa_STN} Total intensity dependence on noise parameters: mean detuning $\overline{\Delta}$ and switching rate $\beta$.}
\end{figure}

We start our analysis by discussing the total scattering intensity $I_{\mathrm{tot}}$, depending on the spectral position of the laser. The full RF intensity as a function of the laser detuning and noise switching rate in the case of symmetric noise is shown in Fig.~\ref{fig:mapa_STN}, where we plot the total scattering intensity obtained by numerical evaluation of Eq.~(\ref{eq:I_tot_ogolnie}). At low switching rates (slow noise), the scattering intensity is the largest when the laser is tuned to one of the two randomly alternating spectral positions of the optical transition, while at high switching rates (fast noise) the strongest scattering occurs at the average value of the transition energy. 

The slow-noise case is easily understood as the quasi-static limit of the random dynamics: over times much longer than the characteristic time scale of the system evolution, $1/\Delta_0$, the system transition energy remains constant at one of the two spectral positions, hence tuning the laser to one of these energies leads to resonant scattering. The dependence on the detuning in this limiting case can be found by setting $\beta=0$, i.e. $B = i[\overline{\Delta}+\mathrm{diag}(-\Delta_0/2,\Delta_0/2)]$, which immediately yields 
\begin{equation}
I_{\mathrm{tot}}^{(\beta=0)}\left(\overline{\Delta}\right) = \frac{1}{2}I_0 \left[ \frac{(\gamma/2)^2}{\left(\overline{\Delta}+\Delta_0/2\right)^2 + (\gamma/2)^2}+\frac{(\gamma/2)^2}{\left(\overline{\Delta}-\Delta_0/2\right)^2 + (\gamma/2)^2} \right]. \label{eq:I_tot_beta0}
\end{equation}
Thus, in the limit of slow noise the scattering intensity reaches half of the noise-free resonant scattering intensity at each of the two resonant spectral positions. 

In the opposite limit of fast noise the random switching takes place many times during the characteristic time $1/\Delta_0$, hence the accumulated dynamical phase slip with respect to the laser light depends only on the averaged transition energy, leading to the resonance condition at $\overline{\Delta} = 0$. By directly evaluating Eq.~(\ref{eq:I_tot_ogolnie}) in the limit of $\beta\to\infty$ one finds in this case
\begin{equation}I_{\mathrm{tot}}^{^{(\beta\to\infty)}}\left(\overline{\Delta}\right) =
I_0\frac{(\gamma/2)^2}{\overline{\Delta}^2 + (\gamma/2)^2},\label{eq:I_tot_beta_infty}
\end{equation}
hence the full standard intensity is recovered at the average spectral position. 
For moderate values of $\beta$ one can neglect $\gamma$ in Eq.~\eqref{eq:I_tot_ogolnie}, hence the noise speed at which the transition between the slow and fast regimes takes place can only depend on $\Delta_0$. Indeed, the scattering intensities at $\overline{\Delta}=\Delta_0/2$ and $\overline{\Delta}=0$ become equal for $\beta/\Delta_0=1/4+O(\gamma/\Delta_0)$.

\begin{figure}[tb]
\includegraphics[width=0.63\linewidth,center]{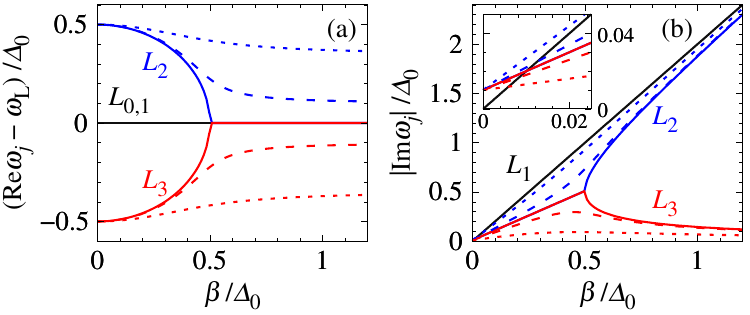}
\caption{\label{fig:bieguny} Positions of the spectral lines (a) and their widths (b) as a function of the  switching rate in the case of symmetric noise, $x=0$ (solid lines), as well as for the asymmetric noise with $x=0.2$ (dashed lines) and 0.7 (dotted lines). Panel (a) corresponds to $\overline{\Delta}=0$. The inset in (b) shows in detail the line widths at very slow switching ($\beta\to 0$).}
\end{figure}

The particular form of the RF spectrum at a given spectral position of the laser is determined by the poles of Eq.~\eqref{eq:widmo_RF_TN}, which we denote by $\omega_j$, $j=0,\ldots,3$, with
\begin{subequations}
\begin{align}
\omega_0 &=\omega_{\mathrm{L}}+i\Gamma, \label{eq:bieguny0} \\
\omega_1 &=\omega_{\mathrm{L}}+i\left(\Gamma+2\beta\right),\label{eq:bieguny1}\\
\omega_{2/3} &=  \omega_{\mathrm{L}} - \overline{\Delta} - i\left(\frac{\gamma}{2}+\beta+\Gamma\right) \pm \frac{\Delta_0}{2}\sqrt{1-\frac{4\beta^2}{\Delta_0^2}+i\frac{4\beta x}{\Delta_0}}. \label{eq:bieguny23}
\end{align}
\end{subequations}
Correspondingly, the Lorentzian features located at the spectral positions $\re\omega_j$ will be labeled by $L_0$, $\ldots$, $L_3$, respectively.
Both $\omega_0-\omega_\mathrm{L}$ and $\omega_1-\omega_\mathrm{L}$ are purely imaginary, corresponding to resonant peaks.

The positions of the spectral features ($\re\omega_j$) as a function of the switching rate $\beta$, calculated from fEqs.~(\ref{eq:bieguny0})--(\ref{eq:bieguny23}), are shown in Fig.~\ref{fig:bieguny}(a) for $\overline{\Delta}=0$. Currently, we focus on symmetric switching, $x=0$, represented by solid lines. Apart from the resonant (central) features $L_0$ and $L_1$ the system in the slow switching limit shows two side peaks $L_2$ and $L_3$. Their positions evolve from $\omega_{\mathrm{L}}\pm\Delta_0/2$ (the locations of the transition energy) in the quasi-static limit towards $\omega_{\mathrm{L}}$, undergoing a qualitative transition at $\beta=\Delta_0/2$, where the characteristic frequencies collapse to a single value of $\omega = \omega_{\mathrm{L}}$. \newtext{Hence, at $\beta=\Delta_0/2$ the RF spectrum changes its form from three lines to a single line.}
This resembles the properties of a damped harmonic system. Here, however, the transition is driven by the switching rate of the noise instead of the damping magnitude, with the cases of slow and fast noise corresponding to the underdamped and overdamped regimes, respectively. According to Eqs.~(\ref{eq:bieguny0})--(\ref{eq:bieguny23}), changing the laser detuning does not affect the spectral positions of the lines $L_2$ and $L_3$ (with respect to the fixed transition energy), while the lines $L_0$ and $L_1$ follow the frequency of the laser.
The broadening of the spectral features, $|\im\omega_j|$, is presented in in Fig.~\ref{fig:bieguny}(b), where we set the instrumental broadening $\Gamma=0$, keep the color coding from Fig.~\ref{fig:bieguny}(a), and omit the $L_0$ peak that has zero width. It follows from 
Eqs.~(\ref{eq:bieguny0})--(\ref{eq:bieguny23}) that the widths are independent of the mean  detuning $\overline{\Delta}$. As follows directly from Eq.~\eqref{eq:bieguny1}, the resonant peak is broadened by $2\beta$ and becomes unbroadened (corresponding to purely elastic scattering) in the quasi-static limit. The side peaks are symmetric in the slow switching regime, with the broadening decreasing to $\gamma/2$ in the quasi-static limit. In the fast regime, when they take the same spectral location, one of them is narrowing asymptotically as $\gamma/2+O[(\Delta_0/\beta)^2]$, while the other one is broadening asymptotically as $2\beta+\gamma/2$.

\begin{figure*}[tb]
\includegraphics[width=\linewidth,center]{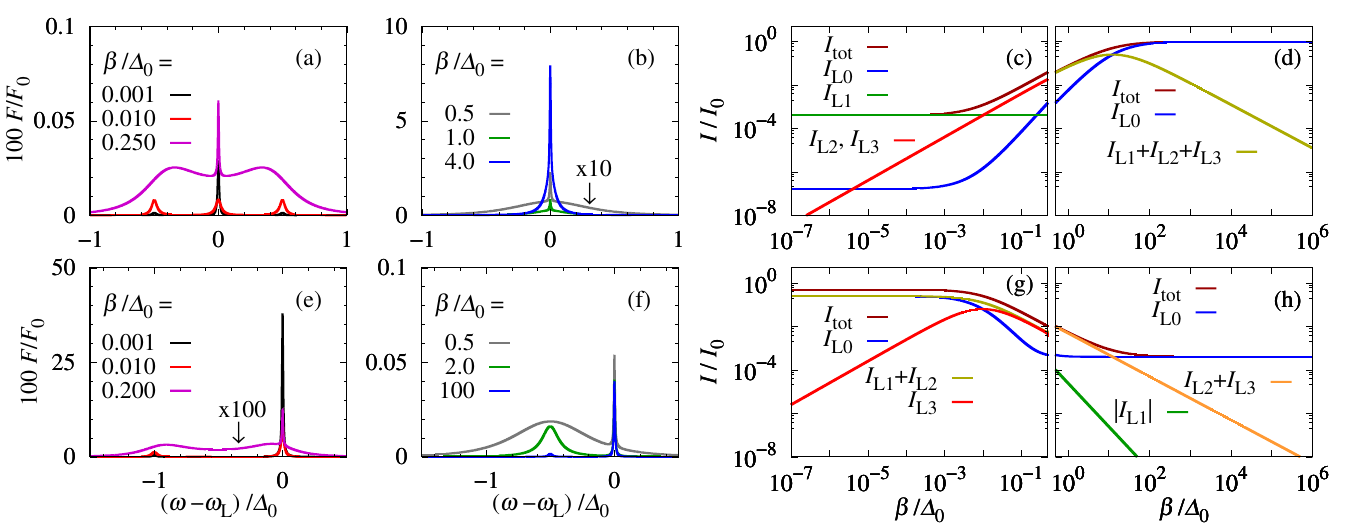}
\caption{\label{fig:STN} (a,b) and (e,f): The RF spectrum for a single-source random telegraph noise, for the laser tuned to the averaged transition energy ($\overline{\Delta}=0$) (a,b) and to the upper position of the transition energy $\overline{\Delta}=\Delta_0/2$ (e,f). Panels (a,e) and (b,f) group the spectra for the slow and fast switching regime, respectively. (c,d) and (g,h): The intensity of the particular components of the RF spectra for the two excitation conditions, respectively, split into the slow (c,g) and fast (d,h) switching case. In the intensity plots, we group certain lines \newtext{and show only their total intensity in some cases, as explained in the text}.}
\end{figure*}
i
Obviously, the peak positions and widths do not provide the complete information about the spectrum, as long as the intensities are not known. In order to fully analyze the spectra we evaluate $F(\omega)$ from Eq.~(\ref{eq:widmo_RF_TN_do_wykresu}) and plot the result in Fig.~\ref{fig:STN}. We start the discussion with the excitation frequency located symmetrically, mid-way between the two positions of the fluctuating transition energy, i.e., $\overline{\Delta}=0$. The spectra for this case, 
for a few selected values of the switching rate $\beta$, 
are shown in Fig.~\ref{fig:STN}(a) and Fig.~\ref{fig:STN}(b) for slow and fast switching, respectively. 
gAs the switching rate grows, the form of the spectrum evolves from a single narrow line, via a triplet of broadened lines that eventually merge into a single line that subsequently narrows down. The quasi-static limit of $\beta\to 0$ (Fig.~\ref{fig:STN}(a), black line) corresponds to the standard result for low-excitation resonance fluorescence from a two-level system, where the scattering spectrum consists exclusively of one narrow line at the spectral position of the laser \cite{ScullyZubairy1997}. Since the laser is detuned from both positions of the transition energy, the overall intensity is very weak. The subsequent evolution of the spectrum is consistent with the structure of the poles discussed above. The appearance of a single line in the fast switching regime can be interpreted again in terms of the averaging of the transition energy on time scales shorter than $1/\Delta_0$ and is therefore the counterpart of the effect observed in Fig.~\ref{fig:mapa_STN}. As the averaged energy level is resonant with the laser, the scattering intensity considerably grows. The line width reduction when speeding up the noise dynamics follows from self-averaging of the fluctuations and is an example of the motional narrowing effect, by analogy to the narrowing of the nuclear magnetic resonance line for a particle that travels very fast through regions of spatially inhomogeneous magnetic field \cite{Bloembergen1948,Berthelot2006}. 

\begin{table}[tb]
\begin{center}
\begin{tabular}{|c|c|c|c|c|}
\hline
& \multicolumn{2}{c|}{$\overline{\Delta}=0$} & \multicolumn{2}{c|}{$\overline{\Delta}=\Delta_0/2$} \\
\hline
 & $\beta\to 0$ & $\beta\to \infty$ & $\beta\to 0$ & $\beta\to \infty$ \\
\noalign{\hrule height 0.75pt}
$I_{\mathrm{tot}}/I_0$ & $\frac{\gamma^2}{\gamma^2+\Delta_0^2}$ & $1$ 
& $\frac{\gamma^2+2\Delta_0^2}{\gamma^2+4\Delta_0^2}$ & $\frac{\gamma^2}{\gamma^2+\Delta_0^2}$  \\
$I_{\mathrm{L0}}/I_0$ & $\frac{\gamma^4}{\left(\gamma^2+\Delta_0^2\right)^2}$ & $1$ 
& $\frac{\gamma^2+\Delta_0^2}{\gamma^2+4\Delta_0^2}$ & $\frac{\gamma^2}{\gamma^2+\Delta_0^2}$\\
$I_{\mathrm{L1}}/I_0$ & $\frac{\gamma^2\Delta_0^2}{\left(\gamma^2+\Delta_0^2\right)^2}$ & $0$ 
& $\frac{\Delta_0^2}{\gamma^2+4\Delta_0^2}$ & $0$\\
$I_{\mathrm{L2}}/I_0$ & $0$ & $0$ & $0$ & $0$\\
$I_{\mathrm{L3}}/I_0$ & $0$ & $0$ & $0$ & $0$ \\
\hline
\end{tabular}
\caption{The limiting values of the intensities of the RF spectrum components in the static and ultrafast limit, for two spectral positions of the exciting laser.}
\label{table:1}
\end{center}
\end{table}

A quantitative understanding of the spectra is possible by combining the information on peak positions and widths, presented in Fig.~\ref{fig:bieguny}, with the peak intensities. The latter are extracted directly by evaluating the pre-factors of the Lorentzian terms in Eq.~\eqref{eq:widmo_RF_TN} and are shown, for $\overline{\Delta}=0$, as a function of the switching rate in Fig.~\ref{fig:STN}(c,d), where we split the result into the slow and fast noise regimes ($\beta<\Delta_0/2$ and $\beta>\Delta_0/2$)\delete{, grouping the intensities of certain lines for clarity (the separate intensities are not necessarily positive)}.
\newtext{
Although the spectrum is always positive, its decomposition into individual peaks is to some extent artificial and some of the components defined in this way may have negative amplitudes if the peaks overlap. Therefore, in some cases we group a few lines that have the same position or the same physical nature and show only the sum of their intensities, so that the presented intensities are positive. 
}
For $\beta\ll\Delta_0$ the total scattering intensity is dominated by the nominally broadened contribution $L_1$. However, as discussed above, in the limit of $\beta\to 0$ its  width decreases to zero, hence the fully elastic scattering is recovered in the static limit.
On the other hand, for $\beta\gg\Delta_0$ total intensity reaches the value characteristic of resonant scattering (Fig.~\ref{fig:STN}(d), purple line) and is dominated by the elastic contribution $L_0$ (blue line), which is consistent with the resonance with the averaged transition energy, leading again to the situation known from a two level system at resonance \cite{ScullyZubairy1997}.
The exact limiting values of the intensities of all the spectral components are collected in Tab.~\ref{table:1}.

We now turn to the case when the laser is tuned to one of the two possible transition frequencies,  $\overline{\Delta} = \Delta_0/2$. 
Fig.~\ref{fig:STN}(e) and Fig.~\ref{fig:STN}(f) show the RF spectrum in this case for $\beta<\Delta_0/2$ and $\beta \ge \Delta_0/2$, respectively. In the quasi-static regime we again observe a single sharp line at laser frequency, but now the intensity is much larger than for $\overline{\Delta}=0$ (Fig.~\ref{fig:STN}(e), black line). As the switching rate grows, this line is accompanied by two broadened lines, that initially appear around the transition energies (one of which now coincides with the laser frequency) and then merge around the central spectral position to disappear again for $\beta\gg\Delta_0$ (Fig.~\ref{fig:STN}(f), blue line) The position of the broadened features is the same as in the previous case (Fig.~\ref{fig:STN}(a,b)) with respect to the transition energies and the position of the sharp peak follows the laser frequency, while the overall intensity now decreases as the switching rate grows.
In this case the laser is tuned to resonance with one of the transition energies, leading to strong scattering in the quasi-static case, which again reproduces the known result for resonant light scattering \cite{ScullyZubairy1997}. On the other hand, in the fast-switching regime, the averaged transition energy is detuned from the laser, hence in this motionally narrowed limit the spectrum corresponds to resonance fluorescence with strongly detuned excitation, showing a weak narrow line at the laser frequency (Fig.~\ref{fig:STN}(f), blue line). 

A quantitative analysis of the intensity of the spectral features contributing to the RF spectrum for $\overline{\Delta}=\Delta_0/2$ is shown in Fig.~\ref{fig:STN}(g) ($\beta<\Delta_0/2$) and Fig.~\ref{fig:STN}(h) ($\beta\ge\Delta_0/2$). In the slow switching regime, only the spectral lines at the laser position contribute, with $L_2$ of negligible intensity (see Tab.~\ref{table:1}) and $L_1$ becoming narrow, as discussed previously. Hence, the elastic scattering fully dominates, as expected for the quasi-static limit. However, the total RF intensity is now lower than the standard resonant scattering intensity $I_0$ (roughly by half), because the probability that the laser is resonant to the transition is now only $50\%$. For fast switching the intensities are consistent with the concept of detuned averaged transition energy, with elastic light scattering ($L_0$ contribution) dominating (Fig.~\ref{fig:STN}h, blue line) and low total intensity.

\subsubsection{\textbf{Asymmetric switching}}

\begin{figure}[tb]
\includegraphics[width=0.60\linewidth,center]{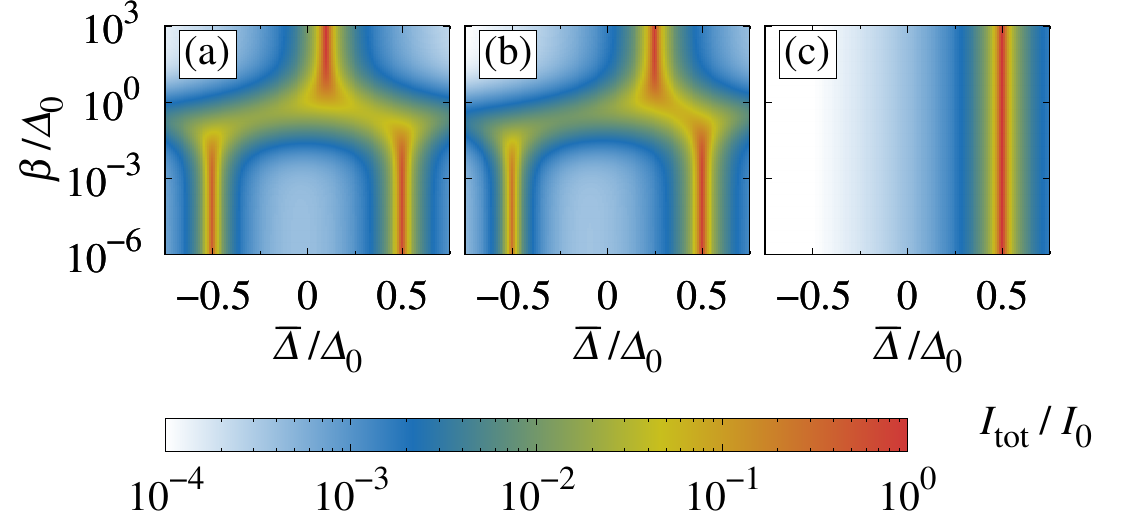}
\caption{\label{fig:mapy_ATN_beta_Delta_mean} \newtext{The total fluorescence intensity as a function of the average detuning and switching rate for a range of values of the degree of asymmetry of the noise: $x = 0.2$ (a), $0.5$ (b) and $1$ (c).}}
\end{figure}

\begin{figure}[tb]
\includegraphics[width=0.60\linewidth,center]{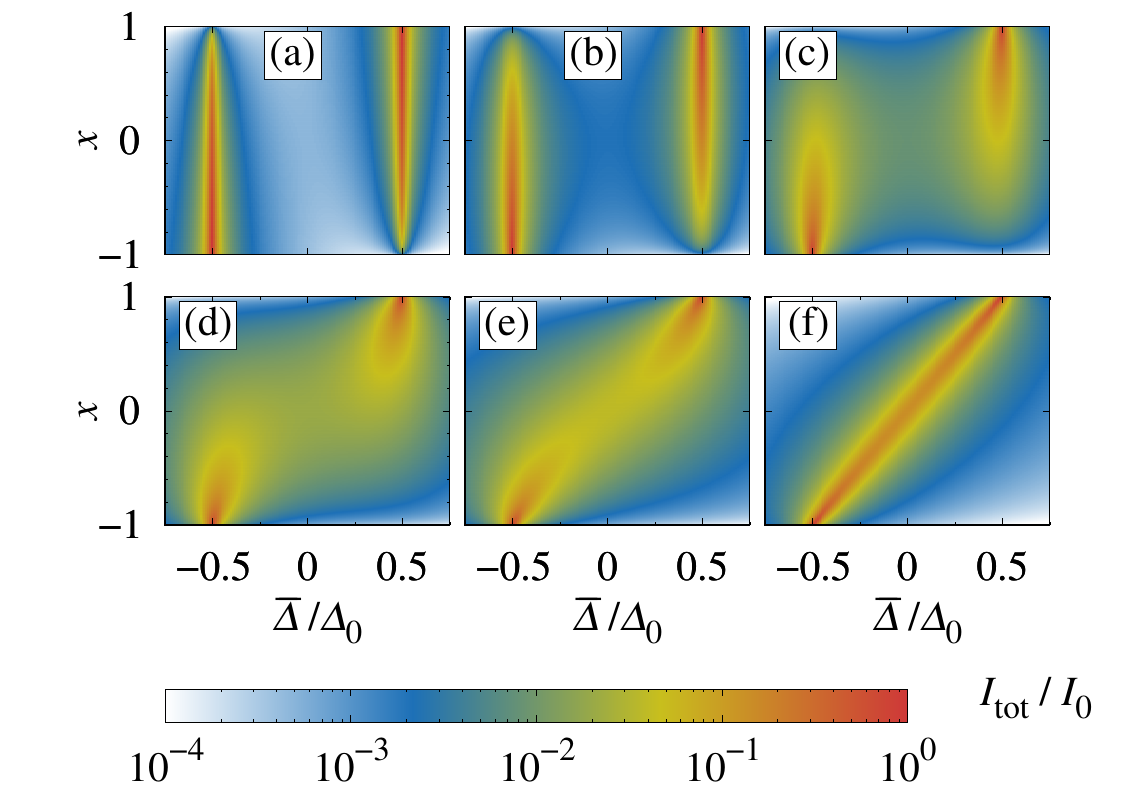}
\caption{\label{fig:mapy_ATN} The total fluorescence intensity as a function of the average detuning and the degree of asymmetry of the noise for a range of values of the total switching rate: $\beta / \Delta_0 = 0.001$ (a), $0.02$ (b), $0.1$ (c), $0.25$ (d), $0.5$ (e) and $2$ (f).}
\end{figure}

In this section we extend our considerations to the case of asymmetric switching, that is, $\beta_\uparrow\neq \beta_\downarrow$, or $x\neq 0$.
At the beginning we discuss the total scattering intensity $I_{\mathrm{tot}}$, \newtext{obtained numerically from Eq.~(\ref{eq:I_tot_ogolnie}) and} now depending  on the spectral position of the laser and on the degree of noise asymmetry, $x=\left(\beta_\downarrow-\beta_\uparrow\right)/\left(\beta_\downarrow+\beta_\uparrow\right)$. 
\newtext{Fig.~\ref{fig:mapy_ATN_beta_Delta_mean}, analogous to Fig.~\ref{fig:mapa_STN},  shows the impact of the asymmetry of the noise. As the preference for the upper position of the transition energy grows with increasing asymmetry, the spectrum gradually evolves into a single line at this spectral position. For slow noise, this happens via transferring the intensity to the right line, without changing the line positions. For fast noise, the position of the line shifts to the right without changing the intensity.}
\newtext{As a complementary view on the same parameter dependence,} Fig.~\ref{fig:mapy_ATN} presents $I_{\mathrm{tot}}$\delete{, obtained numerically from Eq.~(\ref{eq:I_tot_ogolnie}),} as a function of $\overline{\Delta}$ and $x$ for several values of $\beta$. At low switching rates (Fig.~\ref{fig:mapy_ATN}(a,b)), the areas of high RF intensity extend around $\overline{\Delta}=\pm\Delta_0/2$, i.e., when the laser is tuned to one of the two randomly alternating spectral positions of the optical transition. As $\beta$ increases (Fig.~\ref{fig:mapy_ATN}(c-e)), high intensity areas merge, forming finally one  diagonal line (Fig.~\ref{fig:mapy_ATN}(f)).  
The intensity in the slow switching regime is a consequence of the quasi-static dynamics, with the two spectral positions of the transition occurring with the probabilities $p_\pm = (1\pm x)/2$. Indeed, Eq.~(\ref{eq:I_tot_beta0}) is generalized in this case to
\begin{equation}
I_{\mathrm{tot}}^{(\beta=0)}\left(\overline{\Delta}\right) = 
I_0 \left[ p_- \frac{(\gamma/2)^2}{\left(\overline{\Delta}+\Delta_0/2\right)^2 + (\gamma/2)^2} + p_+\frac{(\gamma/2)^2}{\left(\overline{\Delta}-\Delta_0/2\right)^2 + (\gamma/2)^2} \right] . \nonumber
\end{equation}
In the opposite limit of fast noise, the resonance appears at the averaged transition energy, where the average is now weighted by the probabilities $p_\pm$, leading to the averaged energy level of $x\Delta_0/2$. The total intensity in this limit is given by 
\begin{equation*}
I_{\mathrm{tot}}^{^{(\beta\to\infty)}}\left(\overline{\Delta}\right) =
I_0\frac{(\gamma/2)^2}{\left(\overline{\Delta}-x\Delta_0/2\right)^2 + (\gamma/2)^2}, 
\end{equation*}
with $I_{\mathrm{tot}}$ reaching its maximum for $\overline{\Delta}=x\Delta_0/2$.

We next analyze how the positions and widths of the spectral features change with noise asymmetry, parametrized by the parameter $x$. As follows from Eq.~(\ref{eq:bieguny0}) and Eq.~(\ref{eq:bieguny1}), the positions and widths of the two peaks $L_0$ and $L_1$, located at the laser frequency, are independent of the asymmetry. The other two peaks
are represented in Fig.~\ref{fig:bieguny} with dashed and dotted lines for two values of the asymmetry parameter $x$. The spectral positions of these lines (Fig.~\ref{fig:bieguny}(a)) are again bound to the actual spectral positions of the transition at slow switching and converge towards the average frequency as the switching rate grows. However, contrary to the case of symmetric switching, they do not overlap completely but remain separated by a splitting proportional to the asymmetry parameter $x$. Indeed, from Eq.~(\ref{eq:bieguny23}) one finds for $\beta\gg\Delta_0$ the peak positions $\omega_{2,3} = \omega_L - \overline{\Delta} \pm |x|\Delta_0/2$.
The widths of the peaks $L_2$ and $L_3$ are shown with dashed and dotted lines in Fig.~\ref{fig:bieguny}(b) with $\Gamma=0$. The asymptotics of the width of the peaks $L_2$ and $L_3$ for very slow and very fast noise is the same as in the symmetric case but the behavior at intermediate switching rates is different. For small asymmetry of the noise (dashed lines) the widths evolve with $\beta$ in a way similar to the symmetric case. As $x$ grows (dotted lines), the picture changes considerably and one of these lines attains the broadening close to $2\beta$, while the other remains narrow in the whole range of $\beta$. 

\begin{figure}[tb]
\includegraphics[width=0.63\linewidth,center]{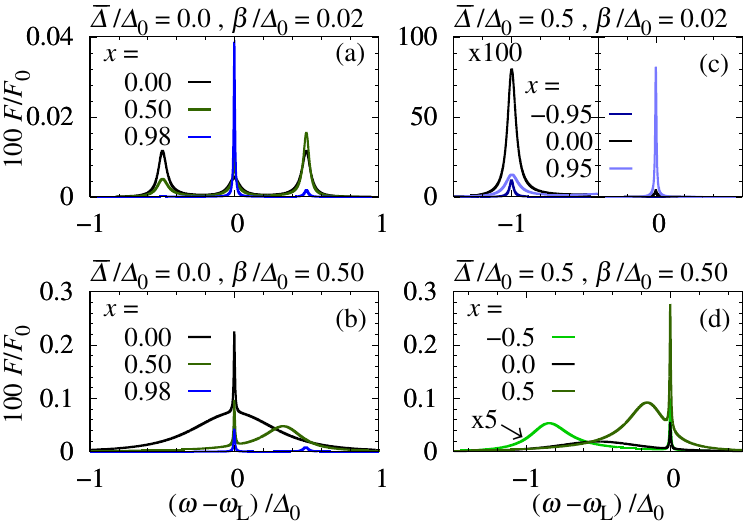}
\caption{\label{fig:ATN_widmo} The dependence of the RF spectrum on the asymmetry of the noise. Panels (a) and (b) show the spectra for $\overline{\Delta}/\Delta_0=0$ in the case of slow noise, $\beta/\Delta_0=0.02$, and moderately fast noise, $\beta/\Delta_0=0.5$, respectively. Panels (c) and (d) present the spectra for $\overline{\Delta}=\Delta_0/2$, for the same two noise rates. The lines on each panel correspond to various values of the asymmetry parameter $x$ as shown.}
\end{figure}

Fig.~\ref{fig:ATN_widmo} presents RF spectra for various values of the laser detuning $\overline{\Delta}$, noise switching rate $\beta$, and noise asymmetry $x$. Each panel corresponds to a certain choice of $\overline{\Delta}$ and $\beta$, and compares the spectrum obtained in the presence of symmetric noise (black lines) with spectra at asymmetric noise (green and blue lines), showing how the intensities and positions of the spectral features evolve with asymmetry. In general, as the asymmetry grows to the limiting values of $x=\pm 1$, the static limit is achieved irrespective of the switching rate $\beta$, so that the spectrum evolves towards a single narrow line at the spectral position of the laser. Figs.~\ref{fig:ATN_widmo}(a,b) show how this happens for the central spectral position of the laser ($\overline{\Delta}=0$). In this case the spectra are mirror-symmetric under the change of the sign of $x$, so only the results for $x>0$ are shown. The initially symmetric spectrum first develops an asymmetry in favor of the most frequently visited spectral position, followed by the decay of the side peaks. As discussed previously, for slow noise, Fig.~\ref{fig:ATN_widmo}(a), the overall scattering intensity in the symmetric case is low, as the excitation is detuned from the two spectral positions of the laser, while for faster noise, Fig.~\ref{fig:ATN_widmo}(b), the intensity is larger as the role of the averaged spectral position increases. However, in the limit of $x=\pm 1$ the noise rate becomes irrelevant and the spectra must converge to the same limit. Hence, the intensity of the central elastic line increases in the former case and decreases in the latter. In Fig.~\ref{fig:ATN_widmo}(c) we show the spectra for the excitation tuned to the upper spectral position of the transition ($\overline{\Delta}=\Delta_0/2$) and for slow noise ($\beta/\Delta_0=0.02$). The spectrum is dominated by the spectral line at the position of the laser (composed of the lines $L_0$, $L_1$ and $L_2$) that gains considerable intensity as $x$ evolves towards $+1$, which means that the excitation becomes resonant with an increasing probability. The other spectral feature (line $L_3$) is enabled dynamically and is always very weak when the noise dynamics is slow (here we magnify it by a factor of 100). It has to vanish at $x\to \pm 1$ and reaches a maximum intensity at $x$ near 0. As can be deduced from Eq.~(\ref{eq:bieguny23}), the position of this line very weakly depends on $x$ when $\beta$ is small. The scattering spectrum for the same spectral position of the laser but faster noise ($\beta/\Delta_0=0.5$) is shown in Fig.~\ref{fig:ATN_widmo}(d). Here not only the intensity but also the position of the off-resonant peak changes, in accordance with Eq.~(\ref{eq:bieguny23}). As follows from Fig.~\ref{fig:mapy_ATN}(e), the overall intensity in this case gradually increases as $x$ changes from $-1$ to $1$, which is reflected in the spectra, both for the resonant and off-resonant peaks. Ultimately, in the static limit of $x\to 1$ (not shown), the intensity of the resonant lines increases by many orders of magnitude and the spectra in Figs.~\ref{fig:ATN_widmo}(c,d) reach the same form of a single, narrow, strong resonant line. 

\begin{figure}[tb]
\includegraphics[width=0.62\linewidth,center]{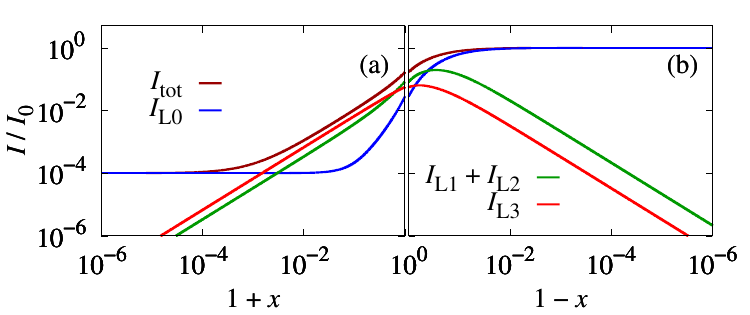}
\caption{\label{fig:ATN_intensywnosc_0,5} RF intensity dependence on $x$ for $\overline{\Delta}=\Delta_0/2$ and $\beta/\Delta_0=0.02$. (a) $x<0$ ($\beta_\uparrow > \beta_\downarrow$);  (b) $x>0$ ($\beta_\uparrow < \beta_\downarrow$).}
\end{figure}

In Fig.~\ref{fig:ATN_intensywnosc_0,5} we analyze quantitatively the intensities of the individual spectral features as a function of the noise asymmetry parameter $x\in [-1,1]$. We restrict our discussion to the excitation tuned to the upper spectral position of the transition, i.e., $\overline{\Delta} = \Delta_0/2$. For the sake of clarity of the presentation, we plot the results for $x<0$ and $x>0$ in a logarithmic scale in Fig.~\ref{fig:ATN_intensywnosc_0,5}(a) and Fig.~\ref{fig:ATN_intensywnosc_0,5}(b), respectively, which reveals power-law dependence as a function of $1-|x|$ as $x$ approaches its limiting values. 
In the quasi-static limit of $x\to \pm 1$, elastic light scattering ($L_0$ spectral line) dominates, as discussed above (blue line in Fig.~\ref{fig:ATN_intensywnosc_0,5}). Obviously the intensity of scattering differs by orders of magnitude in these two limits, as they correspond to strongly detuned and resonant excitation, respectively. In particular, $I_{\mathrm{tot}}\to I_0$ when $x\to 1$ (Fig.~\ref{fig:ATN_intensywnosc_0,5}(b)). In a wide range of intermediate values of noise asymmetry, the noise-induced inelastic scattering dominates (red and green lines in Fig.~\ref{fig:ATN_intensywnosc_0,5}). The inelastic side line ($L_3$, red line in Fig.~\ref{fig:ATN_intensywnosc_0,5}) has its maximum for a slightly asymmetric noise. 

\subsection{N-source telegraph noise}

\begin{figure}[tb]
\includegraphics[width=0.60\linewidth,center]{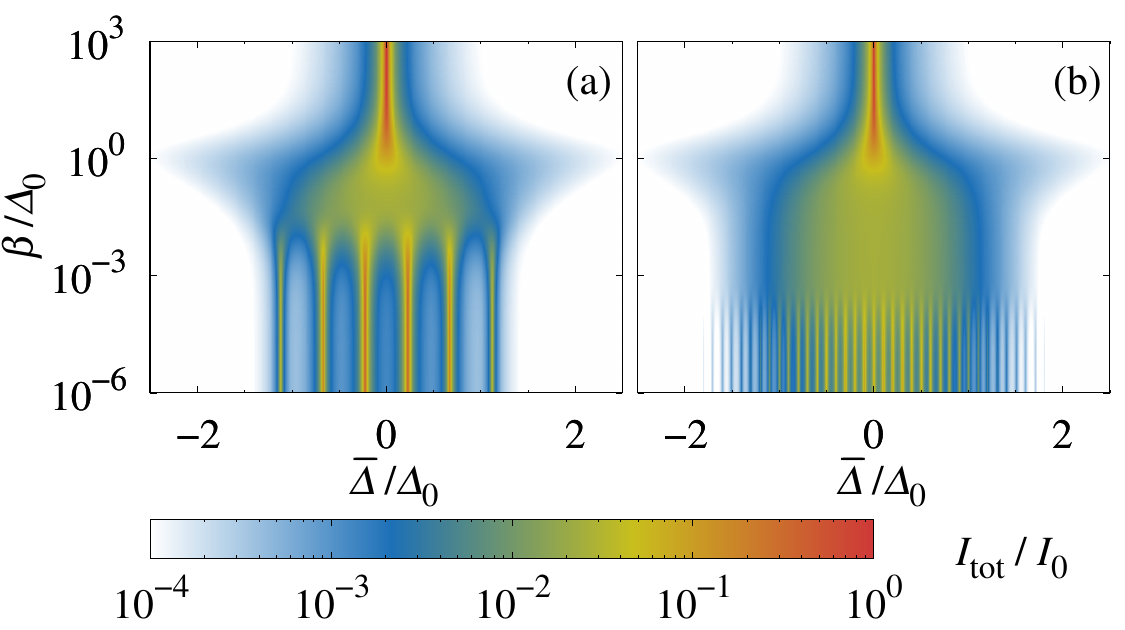}
\caption{\label{fig:mapy_STN_N} Total RF intensity for N identical noise sources as a function of mean detuning for $\beta_\uparrow = \beta_\downarrow = \beta$. (a) $N=5$; (b) $N=100$.}
\end{figure}

In this section we present the results for a system subject to noise originating from $N$ identical additive sources, restricting the discussion to symmetric switching. The total scattering intensity as a function of the spectral position of the laser is shown in Fig.~\ref{fig:mapy_STN_N} for $N=5$ and $N=100$. These results were calculated by numerical evaluation of Eq.~\eqref{eq:I_tot_ogolnie}. In the slow-switching regime, when scanning the laser frequency, we observe $N+1$ resonant frequencies (see Fig.~\ref{fig:mapy_STN_N}(a)). For large $N$, these resonances form a broad spectral feature, with the maximum intensity for the laser tuned centrally ($\overline{\Delta}=0$) (Fig.~\ref{fig:mapy_STN_N}(b)). As the switching rate grows, the resonances merge into a single motionally narrowed line. The appearance of multiple resonances in the slow-noise regime is obviously related to the $N+1$ positions of the transition energies in this case. In the quasi-static limit ($\beta\to 0$), the matrix $B$ becomes diagonal and Eq.~\eqref{eq:I_tot_ogolnie} trivially yields a series of Lorentzian features weighted by the probabilities $p^{(\mathrm{st})}$ that follow the binomial distribution according to Eq.~\eqref{eq:binom}. As a result, the envelope of these resonances forms an approximately Gaussian line (by virtue of the standard Gaussian approximation of the binomial distribution) with a width of $\Delta_0$, which is a consequence of our choice to renormalize the noise amplitudes by $\sqrt{N}$ in Eq.~\eqref{eq:Delta_j}. For $\beta\gg\Delta_0$ the resonances merge, like in the previously discussed case of $N=1$, forming a single narrow Lorentzian resonance at the average transition energy.

\begin{figure}[tb]
\includegraphics[width=0.61\linewidth,center]{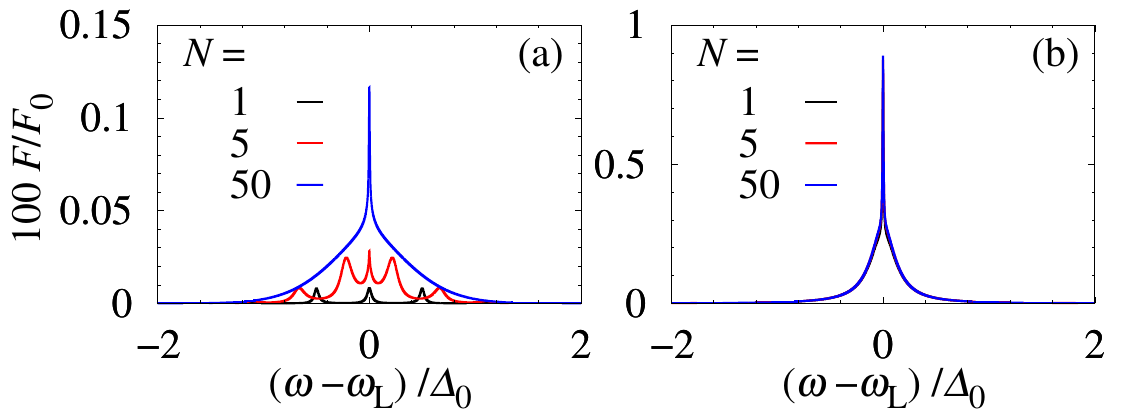}
\caption{\label{fig:STN_N_widmo}  RF spectrum for different numbers of noise sources at $\overline{\Delta} = 0$. (a) $\beta/\Delta_0=0.01$; (b) $\beta/\Delta_0=1$.}
\end{figure}

The RF spectrum for different number of noise sources, calculated numerically from Eq.~(\ref{eq:widmo_RF_TN_do_wykresu}), is shown in Fig.~\ref{fig:STN_N_widmo}. For slow switching and small $N$, the RF spectrum has $N+1$ visible side peaks and the central peak (Fig.~\ref{fig:STN_N_widmo}(a), black and red line). As $N$ increases, the side peaks start to overlap and form a broad feature centered at the laser frequency (blue line in Fig.~\ref{fig:STN_N_widmo}(a)). At $N=50$ the spectrum has reached its asymptotic form and does not change when the number of sources is increased further (which is again due to the normalization of noise amplitudes assumed here). In the fast noise regime, the side peaks are merged into a single feature, as in the previously discussed case of a single source, and there is no visible dependence on $N$ (Fig.~\ref{fig:STN_N_widmo}(b)). The central feature corresponds to the first term in Eq.~\eqref{eq:widmo_RF_TN} and is composed of $N+1$ Lorentzians localized at the laser frequency with widths $2n\beta$, where $n=0, 1, 2, \ldots,N$. In the slow-switching regime, the remaining part of Eq.~(\ref{eq:widmo_RF_TN}) yields $N+1$ Lorentzian
side peaks with an approximately Gaussian envelope at small $\beta$. As in the single-source case, they can be related to the spectral positions of the transition energy resulting from the states of the noise sources. For $\beta \approx \Delta_0$ the spectrum is restructured and all peaks are localized at the laser frequency.

\section{Conclusions}
\label{sec:conclusions}

In this paper we have studied resonance fluorescence from a two-level system subject to classical external noise that leads to fluctuations of the transition energy. We have formulated the general theory of the resonance fluorescence spectrum in the low-excitation regime in the presence of noise that can be described as a stationary Markovian random process on a finite state space, which can approximate an even wider class of Markovian processes. 
We have also described the resonance fluorescence spectrum under uncorrelated noise leading to phase diffusion.
Our theory relates the light scattering spectrum to the formal characteristic of the stochastic noise process.

We have applied our theory to the cases of a single two-state noise source (random telegraph noise) and an arbitrary number of identical sources, where many characteristics can be extracted in an analytical form. Our results show essential differences not only between the phase diffusion and random-telegraph-like processes but also between the regimes of slow and fast dynamics of the random telegraph noise. 
Most remarkably, the resonance fluorescence spectrum changes its form from multiple spectral features or a broad Gaussian feature (depending on the number of noise sources) to a single motionally narrowed line as the noise dynamics gets faster.
In this way, we have demonstrated that light scattering on a two-level system in a noisy environment can yield information on the character of the noise processes experienced by the system. 

These findings may be useful in particular for interpreting experiments on the inherently noisy solid-state systems, where resonance fluorescence finds a constantly growing range of applications. \newtext{In these systems, the typical lifetimes $\gamma^{-1}$ are in the nanosecond range, setting the limit on the field amplitudes for which our low-excitation theory is valid ($\Omega\ll\gamma$). The noise induced by electrical or spin environment \cite{Malein2016,Kasprzak2022,Matthiesen2015}, is typically slow compared to the dynamical time scales of the system. On the other hand, a carefully designed optical experiment \cite{Preuss:22} shows the coexistence of slow (nanosecond time scale) noise with a fast noise component, on picosecond or shorter time scales, which may be due to lattice vibrations. This might open the path to direct verification of our theory. One must note, however, that the transition between the slow and fast regimes in our theory is controlled by the ratio of the noise dynamical rate $\beta$ and noise amplitude $\Delta_0$. The former is an inherent feature of the noisy environment, why the latter may only be modified by engineering the coupling between the emitter and the environment. It may therefore turn out that the most straightforward way to validate the theory would be to use artificially generated mechanical noise, taking advantage of the high flexibility of mechanical signal generation and controllability of the acoustic coupling to solid-state emitters \cite{Weiss2021,Wigger2021}.
}

\section*{Acknowledgments}
This work was supported by the Polish National Science Centre (NCN) under Grant No. 2016/23/G/ST3/04324.

\section*{Data availability}
Any data that support the findings of this study are included within the article.

\appendix
\section{Averaging over the random process}
\label{app:average}

In this Appendix we present the technical details of the averaging in Eq.~\eqref{eq:average}. We start from the basic formula for averaging an arbitrary function of the process state at a finite set of time instants,
\begin{equation} \label{eq:proces_Markova}
\overline{f(\Delta(t_1), ..., \Delta(t_n))} = \sum_{j_1=1}^{k}...\sum_{j_n=1}^{k} p_{j_n,j_{n-1}}(t_n-t_{n-1}) p_{j_2,j_1}(t_2-t_1)p_{j_1}^{(\mathrm{st})} f(\Delta_{j_1},...,\Delta_{j_n}).
\end{equation}
Here we take advantage of the fact that the process is Markovian and stationary, hence the joint probability can be written as a chain of conditional (transition) probabilities with the initial probability distribution assumed to be the stationary probability distribution of the system and the transition probabilities depending only on the time difference.

To find expectation values of exponential terms, $\overline{e^{i\Phi(t_d, t_c)+i\eta\Phi(t_b, t_a)}}$, $\eta=\pm 1$ we divide the time intervals $(t_a,t_b)$  and 
$(t_c,t_d)$ into $N$ and $N'$ pieces, respectively. Then
\begin{equation} \label{eq:czlon_1}
\overline{e^{i\Phi(t_d, t_c)+i\eta\Phi(t_b, t_a)}} = \lim_{N,N'\to\infty}
\overline{e^{i\delta s'\sum_{l=0}^{N'}\Delta(t_l')+i\eta\delta s\sum_{j=1}^{N}\Delta(t_j)}},
\end{equation}
where $N\delta s=t_b-t_a$, $N'\delta s'=t_d-t_c$, $t_j=t_a+\left(j-\frac{1}{2}\right)\delta s$, and $t_l'=t_c+\left(l-\frac{1}{2}\right)\delta s'$.
Upon using Eq.~(\ref{eq:proces_Markova}), one gets
\begin{align*} 
\overline{e^{i\Phi(t_d, t_c)+i\eta\Phi(t_b, t_a)}} = & \lim_{N\to\infty}\lim_{N'\to\infty} \sum_{l_{N'}=1}^{k} ... \sum_{l_1=1}^{k}\sum_{j_{N}=1}^{k} ... \sum_{j_1=1}^{k} q_{l_{N'},l_{N'-1}}^{(1)}(\delta s')...q_{l_2,l_1}^{(1)}(\delta s')e^{i\delta s'x_{l_1}} \\
& \times p_{l_1,j_{N}}(t_c-t_b+\frac{\delta s}{2}+\frac{\delta s'}{2}) q_{j_{N},j_{N-1}}^{(\eta)}(\delta s)...q_{j_2,j_1}^{(\eta)}(\delta s)e^{i\eta \delta sx_{j_1}}p_{j_1}^{(\mathrm{st)}}, \nonumber
\end{align*}
where $q_{l,j}^{(\eta)}(\Delta t)=e^{i\eta x_l\Delta t}p_{l,j}(\Delta t)$. This formula can be rewritten in terms of a product of transition matrices $P_{\tau}$ and $k\times k$ matrices $Q(\tau)$ with matrix elements $q_{l,j}^{(1)}(\tau)$,
\begin{equation} \label{eq:czlon_2}
\overline{e^{i\Phi(t_d, t_c)+i\eta\Phi(t_b, t_a)}} = \lim_{N,N'\to\infty} 
\begin{pmatrix} 1, & \ldots, & 1 \end{pmatrix} \left[Q(\delta s')\right]^{N'-1} e^{i\delta s'x_{l_1'}} P_{t_c-t_b+\frac{\delta s}{2}+\frac{\delta s'}{2}} \left[Q^{*}(\delta s)\right]^{N-1} e^{i\eta\delta sx_{j_1}}p^{(\mathrm{st})}. 
\end{equation}
Keeping in mind that $\delta s \sim 1/N$ and $\delta s' \sim 1/N'$ we have
\begin{equation}
\lim_{N'\to\infty} e^{i\delta s'x_{l_1'}}= 
\lim_{N\rightarrow\infty}e^{i\eta\delta sx_{j_1}} = 1, \label{eq:uproszczenie_1}
\end{equation}
and
\begin{equation} \label{eq:uproszczenie_3}
\lim_{N\rightarrow\infty}\lim_{N'\rightarrow\infty}P_{t_c-t_b+\frac{\delta s}{2}+\frac{\delta s'}{2}} = P_{t_c-t_b}. 
\end{equation}
We write $Q(\delta t) = \mathbb{I} + B\delta t +O(\delta t^2)$.
Only terms up to the linear order survive in the limit $N,N'\to\infty$.
For $\delta t = t/N$ we then obtain
\begin{align}\label{eq:uproszczenie_4}
\MoveEqLeft \lim_{N\to\infty} \left[Q(\delta t)\right]^{N-1} = 
\lim_{N\to\infty} \left(\mathbb{I} + B\frac{t}{N}\right)^{N} = e^{Bt}.
\end{align}
With the use of Eq.~(\ref{eq:uproszczenie_1}),  Eq.~(\ref{eq:uproszczenie_3}), 
and Eq.~(\ref{eq:uproszczenie_4}), Eq.~\eqref{eq:czlon_2} takes the form
\begin{equation*} 
\overline{e^{i\Phi(t_d, t_c)\pm i\Phi(t_b, t_a)}} = \begin{pmatrix} 1 & \ldots & 1 \end{pmatrix} e^{B\left(t_d-t_c\right)} P_{t_c-t_b}e^{B^{(*)}\left(t_b-t_a\right)}p^{(\mathrm{st)}},
\end{equation*}
where the conjugation in the last term corresponds to the '$-$' sign on the left-hand side.


\providecommand{\newblock}{}

\end{document}